\documentclass[11pt,twoside]{article}

\usepackage{asp2006}
\usepackage{graphicx}
\begin{document}
\title{The first stages of the evolution of Globular Clusters}   
\author{Francesca D'Antona, Paolo Ventura}   
\affil{INAF - Osservatorio Astronomico  di Roma, via 
             Frascati 33, 00040 Monte Porzio, Italy}    
\author{Vittoria Caloi}
\affil{INAF - IASF, via Fosso del Cavaliere, 
             00133 Roma, Italy}
\begin{abstract} 
The majority of the inhomogeneities in the chemical composition of 
Globular Cluster (GC) stars appear due 
to primordial enrichment. The most studied model today claims that the ejecta of 
Asymptotic Giant Branch (AGB) stars of high mass ---those evolving during the first 
$\simeq$100Myr of the Clusters life--- directly form a second generation of stars 
with abundance anomalies. In this talk, we review the status of the art 
with regard to this model,
whose major problems are {\it i:} the modelling of the chemical anomalies 
is still not fully complete, and {\it ii:} it requires an IMF peculiarly enhanced 
in the intermediate mass stars. The model predicts enhanced helium abundance in the
stars showing chemical anomalies, and the helium abundance distribution can be
roughly derived from the morphology of the horizontal branch. Such distribution
may possibly help to falsify the model for the first phases of
evolution of GCs. As an illustration, we compare the results of the analysis of the 
HB morphology of some clusters.
\end{abstract}

\section{Introduction}

The observations of GC stars are still to be fully interpreted in a consistent frame. 
Nevertheless, there are a few solid statements which can be put together starting from
observations. In the following we will examine the paradigma of
self--enrichment and point out clearly what is a ``must" and what is only a suggestion.
 
As Raffaele Gratton pointed out in his talk, the latest years have seen
a plethora of beautiful observations of abundances in Globular Cluster stars.
These help to clarify the issue of ``chemical anomalies" that dates back to
the seventies. The anomalies are now observed also at the turnoff (TO) 
and among the subgiants \citep[e.g.,][]{gratton2001, briley2004}, so they must be
attributed to some process of ``self--enrichment" occurring at the first
stages of the cluster life. Therefore, there has been a first epoch 
of star formation which gives origin to the ``normal" (first generation) stars,
that have CNO and other abundances similar to the population II field stars of the
same metallicity. Afterwards, there must have been
some other epoch of star formation, including material 
heavily processed through the CNO cycle.
This material either was entirely ejected by stars belonging to the 
first stellar generation, or it is a mixture of ejected and pristine
matter of the initial star forming cloud. We can derive this statement as a 
consequence of the fact that 
there is no appreciable difference in the metallicity of the ``normal" and
chemically anomalous stars belonging to the same 
GC\footnote{Needless to say, this statement {\it does not}
hold for $\omega$ Cen, which must indeed be considered a small galaxy and not a typical
GC.}. This is an important fact that tells us, e.g., that 
it is highly improbable that the chemical
anomalies are due to mixing of stars born in two different clouds, as there is no reason
why the two clouds should have a unique metallicity. In addition, the clusters showing
chemical anomalies have a huge variety of metallicities, making the suggestion
of mixing of two different clouds even more improbable.  Therefore, the {\bf first statement} 
we can make is: {\it the chemically anomalous
matter which we see in the atmospheres of the chemically anomalous GC stars,  
has been processed by stars belonging to the first stellar generation.} 
This statement does not preclude that the ``first generation" stars,
which we see today, and the progenitors of the CNO processed matter, could have 
been, initially, part of a population much larger than today's GC. This, in fact, 
may be necessary to understand the present number ratios of stars belonging 
to the first and second generation in different GCs.

Presently, there have been interesting attempts to attribute the chemical spreads to 
star formation in the initial proto-GC cloud matter heavily contaminated by
the ejecta of massive stars fastly rotating at velocities close to break up
\citep{decressin}, a view 
which is presented here by Corinne Charbonnel. In this talk we review the status
of the art of the ``classic model", which attributes self--enrichment to a second
star formation phase occurring after the last Supernovae type II exploded in GCs
(the ejecta of SNae are carried easily away from the clusters due to
their high velocity), when the massive Asymptotic Giant Branch (AGB)
stars were evolving. Starting some $\sim 5 \times 10^7$yr from the
birth of the first stellar generation, the massive AGBs 
cycle their envelope material through hot CNO-cycle at the 
bottom of their convective envelopes (Hot Bottom Burning --HBB) and lose them  
in low velocity winds, which may remain inside the cluster and begin a second star
formation epoch.
\begin{figure}
   \centering{
   \includegraphics[width=6.5cm]{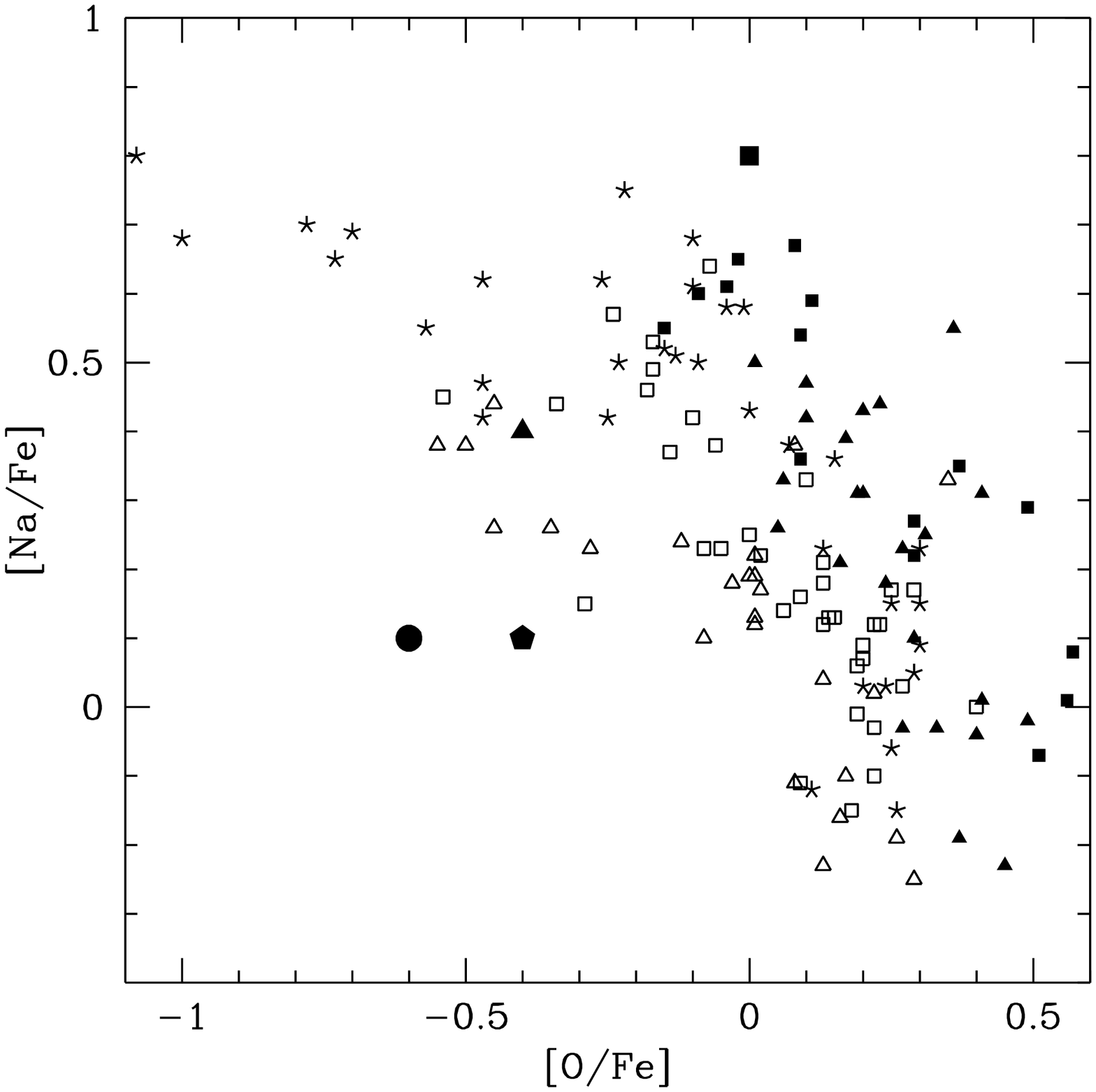}
      \caption{: O-Na anti-correlation for stars in several globular clusters. 
Small symbols indicate stellar abundances (see \cite{ventura-dantona2005b} for details).
[O/Fe] abundances below $\sim$--0.5  can not be explained by selfenrichment by AGB.
The big full (black) symbols represent models for 5M$_\odot$ published in 
\cite{ventura-dantona2006}. The 
	  square is the MLT model, the pentagon and the circle are the
	  standard FST model and the model in which the mass loss rate is reduced by 
	  a factor two, respectively. The triangle is the model in which the lower limit of the
	  cross section of the reaction $^{23}$Na(p,$\alpha)^{20}$Ne is assumed. 
	   }
	 \label{f1}}
   \end{figure}
 
\section{Is the chemistry of massive AGB envelopes resembling the anomalous chemistry in GCs?} 

The HBB that occurs at the bottom of the convective envelopes in
luminous AGB stars is very efficient in a low metallicity environment such as the one in GCs.
While in Population I the main result of HBB is Lithium production and CN cycling ---so
that luminous AGBs lose their Carbon star status---, at low metallicity also the ON cycle
is efficient, and Oxygen is destroyed \citep{ventura2001}. The composition in these
envelopes then depends on the one side on the efficiency of HBB, and on the other side on the
occurrence of the third dredge up, which brings into the envelope the products of nuclear
reactions occurring in the helium shell during the thermal pulses.
The chemistry of anomalous stars in GCs however does not show any sign of processing
in helium burning regions: no indication of s--process enhancement, nor of increased
abundances in the sum of CNO elements \citep{ivans1999, 
briley2002, cohen-melendez2005, cohen2005}. Consequently, either the AGB stars are not the site
of nucleosynthesis of the matter constituting the chemically anomalous stars, or,
we have the {\bf second statement}:
{\it if the AGBs are the source of the chemical anomalies, only
the most massive intermediate mass stars are involved}, those which
process Oxygen to Nitrogen at the bottom of the convective envelope, but have mass loss
rates so high that they pass through few thermal pulses and have only few episodes of the
third dredge up. \cite{ventura-dantona2005} have shown that Oxygen burning
is possible in models that employ a very efficient convection. In turn, the efficient 
convection makes the models more luminous and they lose matter and end their
evolution in a shorter timescale. This
explains the dramatic differences found among the results of different researchers, e.g.
see the comparison between the models in \cite{fenner2004}, that employ low efficiency MLT
convection, vs. the models by \cite{ventura2002}, that employ the 
more efficient convection model by \cite{cgm1996}, as shown in in Figure 7 of 
\cite{ventura-dantona2005b}. Based on the results from models with low convection efficiency,
several researchers have in fact cast severe doubts on the reliability of 
the AGB self-enrichment model.
The agreement between the AGB nucleosynthesis
and the observations is still far from perfect.
The main problem \citep{denis-weiss2001} is to reproduce the
O--Na anticorrelation, but this can be achieved if the cross section
$^{23}$Na(p,$\alpha)^{20}$Ne is taken at the lower limit of the rate by \cite{hale2004}
\citep{ventura-dantona2006}.
Figure \ref{f1} shows the observational data for the anticorrelation and the resulting
abundances in the ejecta of a 5M$_\odot$\ computed with different assumptions on the mass
loss rate and on the cross sections.
Notice that, in any case, the most extreme Oxygen abundances which are found in M13
\citep{sneden2004} reaching [O/Fe]$\simeq-1$, can not be obtained in HBB models: these abundances
require some additional mechanism for Oxygen depletion, for example deep extra--mixing
in giants \citep{denis2006}.

One of the reasons to prefer AGBs as the source of the second generation stars in GCs is
the following. The (scarce) observations of Lithium in
the turnoff stars in NGC 6397 \citep{bonifacio} and NGC 6752 \citep{pasquini}
indicate abundances $\log N(Li)\sim 1.9 - 2.3$, certainly not corresponding
to abundances in a CNO heavily processed environment, {\it unless} this environment
is the HBB AGBs, which indeed may provide such range of abundances,
through the operation of the berillium transport mechanism \citep{cameron}, 
as shown by \cite{ventura2001}.
 
\section{The helium content of the chemically anomalous stars}

A qualitative progress in the understanding of the problem was made when it was shown that
an empiric characteristic of GCs already noticed in the literature 
\citep{catelan1998}, namely the 
presence of abundance anomalies together with peculiar horizontal branch morphologies, 
could be due to
a concomitant helium anomaly in the chemistry. \cite{dantona2002} remarked
a possible interpretation of the morphology of some HBs, very extended in T$_{\rm eff}$,
in terms of an increase in the initial helium content of the cluster stars
which populate the bluer parts of the HB.
There is now a clear correspondence between the chemical peculiarities 
(in particular the O--Na anticorrelation) and the morphology of the horizontal 
branch (see Gratton's talk). Interpreting this latter as an indication that the
helium content of the second generation stars is larger than in the first 
generation, this gives a further element in favour of the AGB self--enrichment
scenario \citep{dantona2002,dantona2004,carretta2006}. 
The proposed model requires that the helium enhancement is present in the whole body of
the stars showing chemical anomalies: consequently, a model in which the chemical anomalies
are due to accretion on already formed stars \citep{dgc1983, suda2006} must be rejected.
The helium yields from the massive AGB ejecta can reach Y$\sim$0.35,
starting from a mere Y=0.24 (the Big Bang abundance). This
result is particularly robust, as it is due primarily to the so called
`second dredge up' phase, which is much less model dependent than the third
dredge up associated with the thermal pulses.
The interpretation for the HB morphology in terms of helium spread received support
from the discovery that the
main sequence of NGC 2808 presents an asymmetric color distribution which can 
best be explained by adding to the normal stars a population of 15--20\% of
stars with very high helium abundance (Y$\sim 40$\%). For
the helium distribution in the stars of
NGC~2808, see the extensive discussion in  \cite{dantona-bellazzini2005}.
The existence of peculiarly blue MSs was also found in $\omega$Cen 
\citep{bedin2004, norris2004, piotto2005}, and for further evidence see Piotto's talk.

If we accept the AGBs as source of the hot--CNO processed material, the helium 
yields expected from these stars \citep{ventura2002} suggest the {\bf third statement}:
{\it the spreads in 
chemical abundances are actually due to the birth of successive generation of 
stars {\sl directly} from the ejecta of the massive AGBs of the first 
generation}. 
Statements two and three suggest another important hint.
There must be a somewhat constrained time for the end of the second 
phase of star formation: it is clear enough that the second star formation stage
must stop abruptly at some epoch. In particular, the
absence of s--process and CNO enhancements in the second generation stars, and
the limitation in Na enhancement point towards a stop in star
formation at the epoch of the evolution of stars having M$< 3.5 -4 $M$_\odot$.
There can be several reasons for this stop, which should be modelled.
For example, the additional energy input by strong UV sources such as the planetary 
nebulae from relatively lower mass progenitors, or by the onset of a 
significant SNIa contribution. 
In any case, stopping early enough the second stage of star formation
contributes to leave a {\it discontinuity} between the helium content 
of the first generation (probably
the Big Bang abundance) and the {\it lowest} helium content of the second
generation. This produces a discontinuity in mass along the red giant
branch, which reflects in a discontinuity in mass along the HB. 
Our {\bf fourth statement} is: {\it if AGBs are the source of the hot--CNO 
processed material, the clusters with chemical anomalies MUST preserve 
a helium discontinuity between the stars of the 
first and second generation}. This is a constraint which is valid until 
we can attribute a cosmological helium content (Y$\sim$0.25) to the first
generation stars. An exception in this respect may be 47 Tuc \citep{salaris1998}.
The presence of a helium discontinuity could falsify the self--enrichment models:
if the anomalies are due to mixing of primordial gas with the gas ejected by
massive stars \citep{decressin}, in fact, there is no reason why a helium discontinuity
should be preserved.

   \begin{figure}
   \centering
\centerline{\hbox{   
   \includegraphics[width=5.5cm]{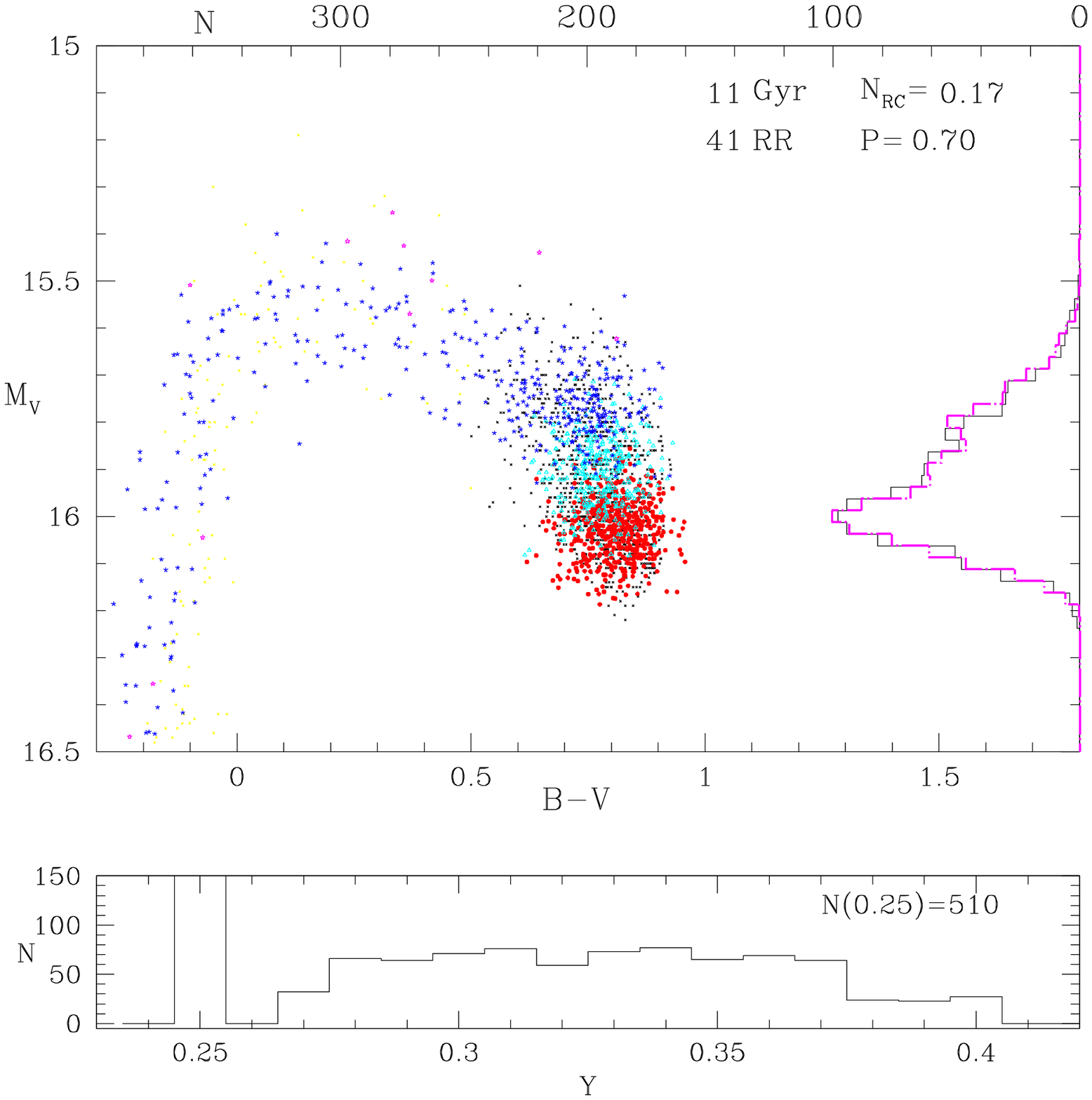}
   \includegraphics[width=5.5cm]{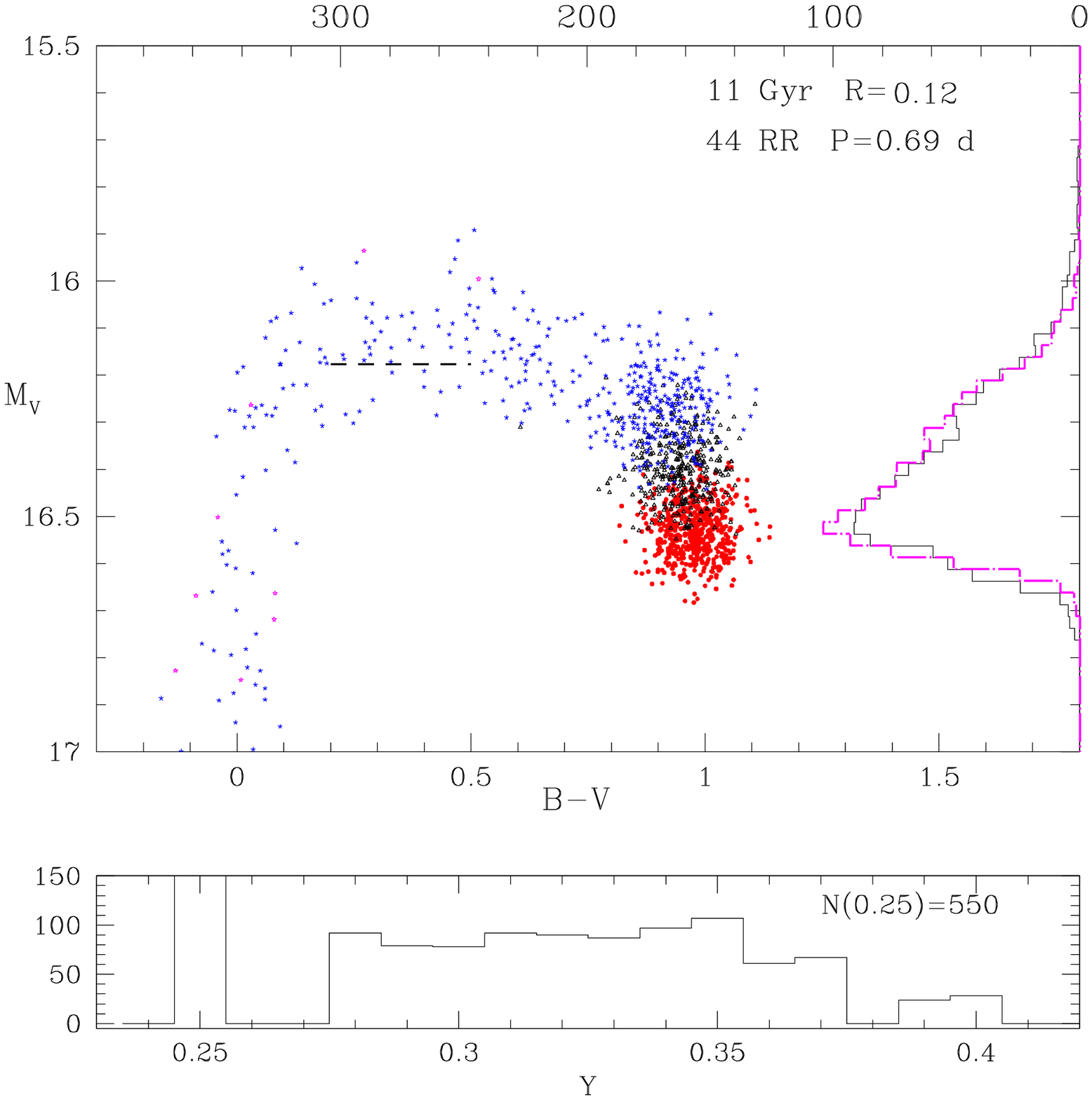}
   }}
      \caption{
	  The fact that the stars in the metal rich clusters
	  NGC 6441 and NGC 6388 have large variations
	  in the helium content is not only testified by the presence of luminous
	  RR Lyr and of an extended blue HB, but also by the luminosity distribution 
	  in the red clump. 
	  This is shown in two synthetic HB simulations for NGC 6441 
	  \citep[][right panel]{caloidantona2006} 
	  and for NGC 6388 (left panel, from Caloi \& D'Antona, in preparation).
	  The different colors refer to stars in different ranges of helium content. The observed
	  clump distributions are shown on the right of each figure, superimposed to the
	  theoretical distribution. The bottom panels show the number vs. helium distributions
      assumed for the simulations. The number of stars with primordial helium Y=0.25,
	  N(0.25), is indicated in the labels.  
	   }
	 \label{f2}
   \end{figure}
The lack of stars in the RR Lyr region in the cluster NGC 2808 can
be considered as an indication in favour of the fourth statement. In fact
this feature suggested us to use the HB ``amplification" to derive information
on the probable helium abundances distribution of the cluster stars \citep{dantona2004}.
We can easily convince ourselves that there are many clusters which also have
a gap in helium content between the two generations, but it is not so evident as in 
NGC~2808 because the different metallicity, mass loss (and possibly also the age) of the cluster 
provide HB masses which cluster either to the blue side of the HB, or into the
red clump, as we will see in the following.   

\section{The helium distribution in metal rich GCs}

The helium spread, although not altering in a significant way the
absolute luminosity of the RR Lyrae in clusters in which there is a
consistent ``first generation" population \citep{dantona2002} produces,
in the particular case of NGC 2808, the small but noticeable difference
in luminosity between the cool side of the blue HB and the hot side of
the red HB \citep{bedin2000}, which, so far, had not been consistently explained. 
\cite{caloidantona2006} recently examined the marked bimodality of very peculiar GCs
NGC 6388 and NGC 6441 \citep{rich-sosin1997}. 
The metallicity of NGC 6441 has been recently confirmed to be high
\citep{clementini2005} and is not consistent with the very long periods of the
RRab variables \citep{pritzl2000} in these two clusters. 
Analysing the HB morphology of NGC~6441 in terms of a helium--rich 
population has provided a coherent interpretation of the three main
features of the anomalous HB of this cluster: 
{\it i:} the long extension in luminosity of the red clump;
{\it ii:}the fact that RR Lyrs have a very long average period, which is unusual 
for a cluster of high metallicity; 
{\it iii: }the extension into the blue of the HB. 

\begin{figure}
   \centering
   \includegraphics[width=6.5cm]{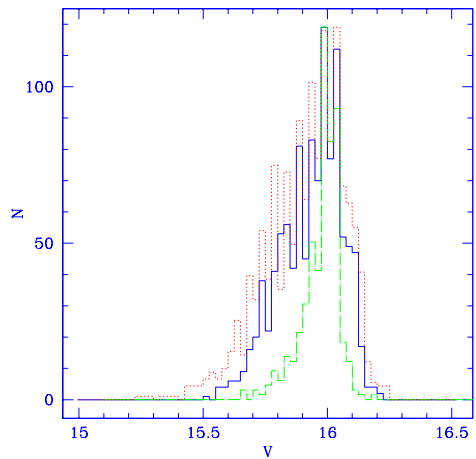}
      \caption{
	  Plot of the observed number of stars versus magnitude for the red horizontal branch
	  of three metal rich clusters: NGC 6441 (dots), NGC 6388 (full line), and 47 Tuc 
	  (dashed line).
	  The magnitudes have been normalized so that the peak in the distribution coincides for 
	  all the clusters. The ``thickness" in magnitude of NGC 6388 and NGC 6441
	  is much larger than in 47 Tuc. The excess of stars at smaller luminosities
	  below the maximum is probably due to the larger observational errors, but the 
	  (asymmetric) excess at
	  higher luminosities is most easily interpreted as due to stars with helium 
	  much higher than normal. 
	   }
	 \label{f3}
   \end{figure}

The synthetic models for the HB of NGC 6441 showed that a helium discontinuity {\it must} be
present (from Y=0.25 to a minimum value of Y$\simeq$0.27, as in the
cluster NGC~2808). Is this a further confirmation of the AGB self--enrichment model?
More precise photometry of the HB would possibly prove a double peak 
in the luminosity distribution of the red clump stars.
Another interesting result of this analysis is the following: 
the fraction of helium rich stars, is much larger than
the fraction of RR Lyr and blue stars (10 -- 12\%). In NGC 6441, 29\%
have Y $>$ 0.33 and $\sim$14\% have Y $>$ 0.35\footnote{This latter figure is
similar to the percentage of stars with Y $\sim$ 0.4 found from 
analysing the MS in NGC 2808 (D'Antona et al. 2005). Thus the 
origin of this tail of very high helium stars, which are not predicted
from AGB model computations, may be similar in these very different GCs.
But these stars represent only the {\it tip of the iceberg} of the self--enrichment
problem}. 
A large fraction of the helium rich stars is contained into the red clump, as we
show in the simulations for NGC 6441 and NGC 6388 in Figure \ref{f2}. In fact, 
the HB lifetime of
metal rich, high helium stars, is spent uniformly in the red and blue parts
of tracks very extended in colours \citep{sweigart-gross1976}. 
Therefore, if there are very luminous, helium rich, RR Lyr, there
must also be very luminous, helium rich, red clump stars, as we see from Figure
\ref{f3}, where we compare the simulated versus observed histogram 
of the number of stars in the red clump vs. the visual magnitude.
\begin{table}[ht]
\caption{Helium history of 4 clusters}
\smallskip
\begin{center}
{\small
\begin{tabular}{|cc|cc|cc|cc|}
\tableline
\noalign{\smallskip}
\multicolumn{2}{c}{NGC 2808}& \multicolumn{2}{c}{NGC 6441}& \multicolumn{2}{c}{NGC 6388}
& \multicolumn{2}{c}{47 Tuc}\\	
  Y	 & \% &  Y	 & \% &  Y	 & \% &  Y	 & \% \\
\noalign{\smallskip}
\tableline  
\noalign{\smallskip}
0.24	& 50 &	0.25      & 38	&.25	& 39	&.25	&75 \\
0.26-0.29 &	35&	0.27-0.35	&48&	0.27-0.35&	41	&0.27-.32	&25 \\
$\sim$0.4 &	15 &	$>$0.35 &14 &	$>$0.35 &	20 && \\		
\noalign{\smallskip}
\tableline
\end{tabular}
}
\end{center}
\label{table1}
\end{table}

Table \ref{table1} shows the preliminary ``helium histories" of four clusters, based  
on synthetic models of their HBs. The analysis has been done also for 47 Tuc. The different
thickness in luminosity of the red clump is the main reference for the analysis
of the three metal rich clusters, while the results for NGC 2808 come from
\cite{dantona-bellazzini2005}. Notice
that the small percentage (25\%) of stars with high helium in 47~Tuc is in contradiction with
the observation that CN strong and CN weak stars in the cluster 
are about in similar percentages \citep{briley2004}. If {\it no helium gap is present},
the percentage can reach $\sim$40\%. An escape from this problem can be found 
if the first stellar generation in
47~Tuc has a larger initial helium content \citep{salaris1998}.

We have to mention also the possibility that some clusters have today only members
with  helium content of larger than the primordial value. This is possibly the
situation in M13, one of the 
classic ``second parameter" pair M3 and M13, according to \cite{caloi2005} 
suggestion. If we take this suggestion seriously, the {\bf fifth statement} is: {\it
the fraction of 1st stellar generation (no chemical anomalies, normal Y) which 
is today present in the cluster can vary among clusters.
The clusters with predominantly blue HB might have lost "almost" all the 1st generation.}
Although very difficult to be accepted, this statement puts the accent on
the fact that the primordial GCs might have been much more massive than today's.
In this respect, the problem of the initial mass function might find a solution, if a large
fraction of the first stellar generation has been lost.

\bibliographystyle{aa}

\acknowledgements 
It is a pleasure to thank the organizers of this Conference in honour of Cesare Chiosi, 
a meeting which has been spectacular both from the scientific and
environmental point of view. Happy birthday, Cesare!

\end{document}